\documentclass[apjl]{aeb_emulateapj}
\usepackage{natbib}
\def\GHz{{\rm GHz}} %.......Gigahertz
\def\m{{\rm m}} %...........meters
\def\cm{{\rm c}\m} %........centimeters
\def\pc{{\rm pc}} %.........parsecs
\def\G{{\rm G}} %...........gauss
\def\muG{\mu\G} %...........microgauss
\def\rad{{\rm rad}} %.......radians

\renewcommand{\d}{{\rm d}}
\newcommand{\e}{{\rm e}}

\newcommand{\RM}{{\rm RM}}

\newcommand\bmath[1] {\mbox{\boldmath$\rm #1$}}

\begin{document}

\title{Signatures of Relativistic Helical Motion in the Rotation Measures of AGN Jets}

\author{
Avery E.~Broderick\altaffilmark{1} \& 
Abraham Loeb\altaffilmark{2}
}
\altaffiltext{1}{Canadian Institute for Theoretical Astrophysics, 60 St.~George St., Toronto, ON M5S 3H8, Canada; aeb@cita.utoronto.ca}
\altaffiltext{2}{Institute for Theory and Computation, Harvard University, Center for Astrophysics, 60 Garden St., Cambridge, MA 02138, USA}

\shorttitle{Relativistic Helical Motion and AGN Jet RM's}
\shortauthors{Broderick \& Loeb}

%\submitted{Submitted version July 16, 2009}

\begin{abstract}
Polarization has proved an invaluable tool for probing magnetic fields
in relativistic jets.  Maps of the intrinsic polarization vectors have
provided the best evidence to date for uniform, toroidally
dominated magnetic fields within jets.  More recently, maps of the
rotation measure ($\RM$) in jets have for the first time probed the field
geometry of the cool, moderately relativistic surrounding material.
In most cases, clear signatures of toroidal magnetic field are
detected, corresponding to gradients in $\RM$ profiles transverse to
the jet.  However, in many objects these profiles also display marked
asymmetries which are difficult to explain in simple helical jet
models.  Furthermore, in some cases the $\RM$ profiles are strongly
frequency and/or time dependent.  Here we show that these features may
be naturally accounted for by including relativistic helical motion in
the jet model.  In particular, we are able to reproduce bent $\RM$
profiles observed in a variety of jets, frequency dependent $\RM$
profile morphologies and even the time dependence of the $\RM$
profiles of knots in 3C 273.  Finally, we predict that some sources
may show reversals in their $\RM$ profiles at sufficiently high
frequencies, depending upon the the ratio of the components of jet
sheath velocity transverse and parallel to the jet.  Thus,
multi-frequency $\RM$ maps promise a novel way in which to probe the
velocity structure of relativistic outflows.
\end{abstract}


\maketitle

\section{Introduction} \label{I}
Polarization observations of relativistic jets have been instrumental
in identifying the presence of large-scale ordered magnetic fields
\citep{Tayl-Perl:92,Tayl:98,List-Mars-Gear:98,Tayl:00,Zava-Tayl:01,Zava-Tayl:02,Zava-Tayl:03,Zava-Tayl:04,Asad_etal:08a,Asad_etal:08b,Khar_etal:09,OSul-Gabu:09}.
For nearly all active galactic nuclei (AGN) jets, Faraday rotation
dominates the polarization signal, scrambling the polarization angles
of the intrinsic emission on angular scales small compared to the jet
width.  This can be ameliorated by determining the rotation measure
($\RM$), via multi-wavelength polarization observations, and then 
removing the associated distortion using the standard relation,
\begin{equation}
\Psi_{\rm int} = \Psi_{\rm obs} - \RM\,\lambda^2\,,
\end{equation}
where $\Psi_{\rm int, obs}$ is the intrinsic/observed polarization
angle.  Typical $\RM$'s for AGN jets and cores range from
$10\,\rad\,\m^{-2}$ to $10^3\,\rad\,\m^{-2}$, though it can reach
$10^5\,\rad\,\m^{-2}$ in some extreme cases.  Thus, at $1\,\GHz$
frequencies, the intervening Faraday rotation results in polarization
angle rotations of roughly $6^\circ$--$600^\circ$.  Nevertheless, once
the Faraday rotation is solved for, via multi-wavelength polarization
observations, and removed, most jets show remarkably uniform
polarization maps, generally aligned with the jet axis.  Both the
polarization fraction and the orientation has been used to argue that
in the observer frame jets are overwhelmingly toroidally dominated
\citep{Lyut-Pari-Gabu:05}.

Observations of the $\RM$ itself provide information about the
intervening magnetized plasma.  There are multiple potential sources
for the $\RM$'s including the Galactic interstellar medium, the
intracluster medium near the AGN and cooler, less relativistic
material surrounding the jet itself.  The Galactic $\RM$ itself can be
on the order of $10^2$--$10^3\,\rad\,\m^{-2}$ and varies considerably on large
scales, presumably as a consequence of the structure of the Galactic
magnetic field \citep{Eker-Lequ-Moff-Seie:69,Men-Ferr-Han:08,Nout_etal:08}.
The magnetic field strength and geometry within the intracluster
medium is poorly constrained, though few sources lie sufficiently near
the centers of cooling core clusters for this to dominate
\citep{Cari-Tayl:02}.  In contrast,
Faraday rotation intrinsic to the jet environment itself is expected
to produce large $\RM$'s.

The morphology of the Faraday rotating medium surrounding AGN jets is
still poorly constrained.  While the lack of Faraday depolarization
implies that the medium must be in the foreground and itself produce
negligible emissions, presently it is not
possible to distinguish between nearby unrelated ionized gas
clouds and rotation within the near side of a jet sheath, presumably
containing an ordered helical magnetic field.  Free-free absorption
from $\pc$-scale foreground clouds has been detected in Cen A and NGC
1275, implying that these clouds exist in at least some sources
\citep{Jone_etal:96,Walk_etal:00}.  In these one would expect
roughly random $\RM$ gradients, associated with the random
orientations of the cloud magnetic fields.  However, in a handful of
cases thus far ordered $\RM$ gradients transverse to the jet axis have
been observed, providing tantalizing evidence for the presence of
helically magnetized sheaths surrounding AGN jets
\citep{Asad_etal:08a,Asad_etal:08b,Khar_etal:09,OSul-Gabu:09}.  It is
this latter possibility that we restrict ourselves here.

For static, axisymmetric jet sheaths, the transverse $\RM$ profiles are
roughly linear and symmetric about the jet axis, up to a uniform
offset depending upon the magnetic pitch angle.  While for many
sources this is sufficient
\citep{Asad_etal:08b,Khar_etal:09,OSul-Gabu:09}, in a number of cases
the $\RM$ profiles display significant asymmetries
\citep{Asad_etal:08b,OSul-Gabu:09}.  Furthermore, these asymmetries
can be both
time and frequency dependent.  While relativistic motion along the jet
can reduce the $\RM$ in the observer frame considerably, it is unable
to produce these asymmetric profiles \citep{Lope:06}.  Here we
consider the effects of relativistic {\em helical} motion.  This is expected
theoretically in both magnetohydrodynamic and force-free jet
models, as well as motivated by jet simulations, though these
typically cannot probe the necessary dynamic range to make direct comparisons
\citep{Vlah-Koni:04,DeVi-Hawl-Krol-Hiro:05,McKi:06,Komi-Bark-Vlah-Koni:07,Tche-McKi-Nara:08,McKi-Blan:09}.
Within the context of a simple helical jet model, it is possible to
reproduce both,  the observed asymmetries and the frequency \&
temporal behavior of jet $\RM$ profiles, without appealing to
asymmetric structures or special viewing angles.

\section{Faraday Rotation in Relativistic Bulk Flows} \label{sec:FRiRM}
Faraday rotation is a result of the different phase velocities of the
two, nearly circularly polarized\footnote{This is generally true only
  for cold, ionic plasmas.  In pair plasmas the eigenmodes are linear,
  while for plasmas with relativistic temperatures or substantial
  non-thermal components, the eigenmodes are typically significantly
  elliptical.  However, the detection of $\RM$ gradients across AGN
  jets provides strong evidence for the presence of a significant, if
  not dominant, ionic component in the plasma surrounding the jet.  For
  this reason, we have ignored the possible presence of a pair plasma
  in our analysis.  Should the jet sheath be comprised of an admixture
  of the two, $n$ simply represents the excess electron
  density \citep{Hall-Shuk:05}.  That the plasma is sufficiently cold
  is generally a good assumption far from internal shocks, and
  observationally supported by the lack of significant Faraday
  depolarization, and hence emission from within the Faraday rotating
  sheath.}, electromagnetic eigenmodes of magnetized plasmas.  The
subsequent accrual of a phase difference results in a local rotation
of the polarization plane.  Critical to this interpretation is the
assumption that the plasma eigenmodes propagate nearly adiabatically.
While changes in the underlying plasma properties (density and
magnetic field) can lead to violations of this condition, for simplicity we will
only consider cases in which the relevant plasma parameters change
sufficiently slowly that these may be ignored
\citep{Enss:03,Brod-Blan:09}.  Furthermore, we will assume that the
plasma smoothly joins the sub-relativistic material surrounding the
jet, and thus will not consider the relativistic aberration of the
resultant polarization.  In this case, locally, the polarization
angle, $\Psi$, changes by an amount identical to the phase difference
accrued, $\Phi$.  That is, in the plasma rest frame
\begin{equation}
\d\Psi = \d\Phi = \frac{16\pi^3 e^3}{m_e^2 c^2 \tilde{\nu}^{2}} n \tilde{\bmath{B}}\cdot\d\tilde{\bmath{\ell}}\,,
\end{equation}
where $n$ and $\tilde{\bmath{B}}$ are the proper electron density and
rest-frame magnetic field, $\tilde{\nu}$ is the rest-frame frequency and
$\d\tilde{\bmath{\ell}}$ is the rest-frame distance element.  Since the
total accrued phase is a Lorentz scalar, and we assume the plasma
smoothly joins a non-relativistic flow, we obtain
\begin{equation}
\Delta \Psi = \frac{16\pi^3 e^3}{m_e^2 c^2} \int
\frac{n}{\tilde{\nu}^{2}} \tilde{\bmath{B}}\cdot\d\tilde{\bmath{\ell}}\,.
\end{equation}

Presently, this integral is performed in the comoving frame of the
plasma.  However, we may use the scalar nature of $\Psi$ to rewrite it
in terms of the observed frequency, $\nu$, and an integral in the
observers frame, $\d\ell$.  To do this, let us begin by defining the
plasma four-velocity\footnote{Henceforth we set $c=1$ unless otherwise
  noted.}, $u^\mu = \gamma(1,\bmath{\beta})$.  In either the
force-free or magnetohydrodynamic prescriptions, the electric field
vanishes in the plasma frame, and thus $F^{\mu\nu} u_\nu = 0$, where
$F^{\mu\nu}$ is the electromagnetic field tensor.  This implies that
$F^{\mu\nu} = \epsilon^{\mu\nu\alpha\beta} u_\alpha b_\beta/2$, where
$b^\mu=(b^t,\bmath{b})$ is the four-vector corresponding to the
magnetic field in the plasma rest frame and orthogonal to $u^\mu$.
Due to the latter condition, the components of $b^\mu$ are not
independent, and in particular $b^t = \bmath{\beta}\cdot\bmath{b}$.
Finally, let us define the electromagnetic wave-vector,
$k^\mu=2\pi\nu(1,\bmath{\hat{k}})$, which is related to the
electromagnetic wave trajectory, $x^\mu(\eta)$, by
$\d x^\mu/\d\eta = k^\mu$.  This implies that the line element, as
measured in the plasma frame is
$
\d\tilde{\ell}^\mu
=
\d x^\mu + u_\nu \d x^\nu u^\mu
=
\left(k^\mu + u_\nu k^\nu u^\mu\right) \d\eta\,,
$
and thus, $\d\tilde{\ell} = - u_\mu k^\mu \d\eta$.  Similarly, in the
lab frame, $\d\ell = k^t\d\eta$.
Therefore,
\begin{equation}
\begin{array}{c}
\displaystyle
\tilde{\nu} = -\frac{u^\mu k_\mu}{2\pi} = \gamma \left(1-\bmath{\beta}\cdot\bmath{\hat{k}}\right)\nu
\\
\\
\displaystyle
\tilde{\bmath{B}}\cdot\d\tilde{\bmath{\ell}} = b_\mu \d\tilde{\ell}^\mu = \left(\bmath{\hat{k}}-\bmath{\beta}\right)\cdot\bmath{b} \,\d\ell
\,.
\end{array}
\end{equation}
Inserting these into the expression for rotation of the polarization
gives,
\begin{equation}
\Delta\Psi
=
\frac{16\pi^3 e^3}{m_e^2 c^2} \frac{1}{\nu^2}
\int
n
\frac{
\left(\bmath{\hat{k}}-\bmath{\beta}\right)\cdot\bmath{b}
}{
\gamma^2\left(1-\bmath{\beta}\cdot\bmath{\hat{k}}\right)^2
}
\d\ell\,.
\end{equation}
The associated rotation measure is then simply
\begin{equation}
\RM
\equiv
\frac{\partial\Delta\Psi}{\partial\lambda^2}
=
0.812\,
\int
n
\frac{
\left(\bmath{\hat{k}}-\bmath{\beta}\right)\cdot\bmath{b}
}{
\gamma^2\left(1-\bmath{\beta}\cdot\bmath{\hat{k}}\right)^2
}
\d\ell
\,\,\,\,\rad\,\m^{-2}\,,
\label{eq:RM}
\end{equation}
where $n$, $b$ and $\ell$ are measured in $\cm^{-3}$, $\muG$ and $\pc$,
respectively.

When $\beta=0$, this rotation measure reduces to the standard
expression, with $n$ and $\bmath{b}$ now corresponding the plasma
density and magnetic field strength in the observer frame.  However,
when $\beta\ne0$, some care must be taken in interpreting these
quantities, as both are defined in the plasma rest frame.  In
particular, the observer frame magnetic field, $\bmath{B}$, is
related to $\bmath{b}$ by
\begin{equation}
\bmath{B}
=
\gamma \left( \bmath{1} - \bmath{\beta}\bmath{\beta}\right)\cdot\bmath{b}
\quad\Leftrightarrow\quad
\bmath{b}
=
\frac{1}{\gamma} \left( \bmath{1} + \gamma^2 \bmath{\beta}\bmath{\beta}\right)\cdot\bmath{B}
\,,
\end{equation}
which may be found by inspecting $F^{\mu\nu}$ directly.  For large
$\gamma$, $\bmath{B}$ will appear dominated by the components
transverse to the plasma motion.

The relativistic motion of the plasma has two important consequences.  The
first is to change the magnitude of the magnetic field along the line
of sight due to relativistic aberration, producing the velocity
dependent term in the numerator of Equation (\ref{eq:RM}).  The second is
simply to Doppler-shift the relevant plasma frequency, signified by the
denominator in Equation (\ref{eq:RM}), resulting in a correspondingly
higher or lower $\RM$ as measured in the observer frame.  This second
effect can produce profound variations in the $\RM$ when
$\bmath{\beta}$ passes nearly parallel to the line of sight.  The
ratio of the $\RM$ of approaching and receding plasma flows is
roughly $4\gamma^2$, which for even moderate Lorentz factors can be
considerable.

\section{Rotation Measure Profiles of Jets} \label{sec:RMPoJ}
Having an expression for the relativistic rotation measure in hand, we
now must specify the properties of the Faraday rotating medium.  For
illustrative purposes we adopt an extraordinarily simplified model:
a cylindrically symmetric plasma sheath surrounding a jet core,
appropriate at large distance from the jet launching region.  While the
cylindrical approximation is poor for viewing angles within the jet
opening angle, for most cases of interest it is sufficient, allowing
us to remove a degree of freedom from the modeling of the Faraday
screen.  The density peaks at some distance from the jet axis, falling
exponentially beyond (our results are qualitatively independent of the
radial density profile).  We assume helical geometries for both the
magnetic field and the plasma four-velocity.  Explicitly, in
cylindrical coordinates, 
\begin{equation}
\begin{array}{c}
\displaystyle
n = n_0 \left(\frac{R}{w_{\rm jet}}\right)^2 \e^{-R^2/w_{\rm jet}^2}\\
\\
\displaystyle
b^z = b_0\cos\alpha_b
\,,\quad
b^R = 0
\,,\quad
b^\phi = \frac{b_0\sin\alpha_b}{R}\\
\\
\displaystyle
u^z = \beta\cos\alpha_u
\,,\quad
u^R = 0
\,,\quad
u^\phi = \frac{\beta\sin\alpha_u}{R}\,,\\
\\
\end{array}
\end{equation}
where $R$ \& $\phi$ are the cylindrical radius and polar angle,
respectively, $w_{\rm jet}$ is the characteristic width of the jet sheath, $n_0$ and
$b_0$ are the density and field strength scaling (and 
are ultimately unimportant to the question of the $\RM$ asymmetry),
$\beta$ is the velocity and $\alpha_u$ and $\alpha_b$ are the velocity
and magnetic field pitch angles, respectively.  Finally, we define the
angle between the line of sight and the jet axis to be $\Theta$, i.e.
$
\hat{k}^z = \cos\Theta
$, $
\hat{k}^R = \sin\Theta \cos\phi
$ and $
\hat{k}^\phi = -\sin\Theta\sin\phi/R
$.
Thus, $\Theta=0^\circ$ and $\Theta=90^\circ$ correspond to lines of
sight along and orthogonal to the jet axis, respectively.  For a
variety of values for these parameters, we compute the $\RM$ as
defined in Equation (\ref{eq:RM}) as a function of transverse
position, thereby constructing profiles across the jet.

\begin{figure}
\begin{center}
\includegraphics[width=0.43\textwidth]{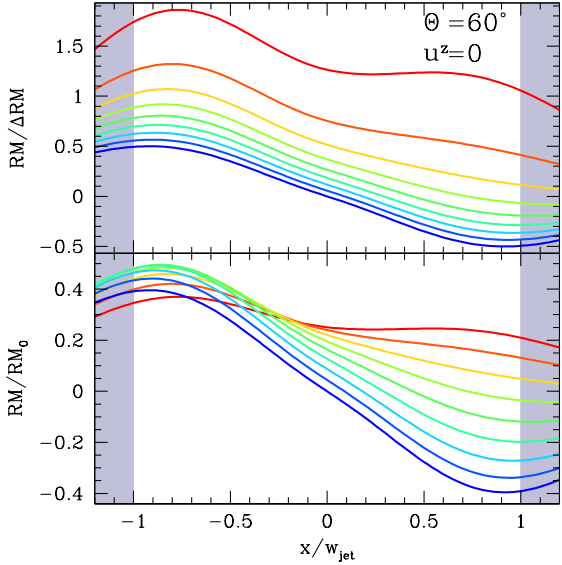}
\end{center}
\caption{Rotation measure profiles of a static jet ($u^z,u^\phi=0$) as viewed
  at $\Theta=60^\circ$, for magnetic pitch angle, $\alpha_b$, ranging
  from $10^\circ$ (red) to $90^\circ$ (blue) in steps of $10^\circ$.
  {\em Top:} The $\RM$ profile normalized such that the variation in
  the rotation measure is unity.  {\em Bottom:} The $\RM$ profiles in units
  of $81.2\,\rad\,\m^{-2}$, corresponding to the $\RM$ generated by
  Faraday screen with electron density $1\,\cm^{-3}$, magnetic field
  $1\,\muG$ and line-of-sight depth $1\,\pc$.  In both panels the
  grayed regions show areas outside the jet core in projection, and
  thus presumably in which the $\RM$ cannot be
  measured.} \label{fig:static}
\end{figure}

The spatial variations in the density and orientation of the magnetic
field necessarily lead to variation in the $\RM$ across the jet.
In the limit of a static jet, where the plasma is at rest in the
observer frame, the $\RM$ profile is shown in Figure \ref{fig:static}
for a variety of magnetic pitch angles, as viewed from
$\Theta=60^\circ$.  For purely toroidal magnetic fields
($\alpha_b=90^\circ$, blue), the $\RM$ profile is symmetric and
approximately linear, only departing from a line at the sheath
boundaries due to the precipitous decrease in the electron density.
In contrast, for magnetic fields nearly
parallel to the jet ($\alpha_b=0^\circ$, red) the background density
variation is clearly imprinted in the $\RM$ profile.  Due solely to
geometry, the degree to which this occurs depends upon $\Theta$,
with the largest deviations at small $\Theta$ and $\alpha_b$.
However, the absolute variation in the $\RM$ across the jet in this
case is considerably reduced, and in the limit of $\alpha_b=0^\circ$
vanishes completely.  More importantly, polarization maps of jets
typically imply $R b^\phi/b^z\gtrsim 1$ \citep{Lyut-Pari-Gabu:05}.

\begin{figure}
\begin{center}
\includegraphics[width=0.43\textwidth]{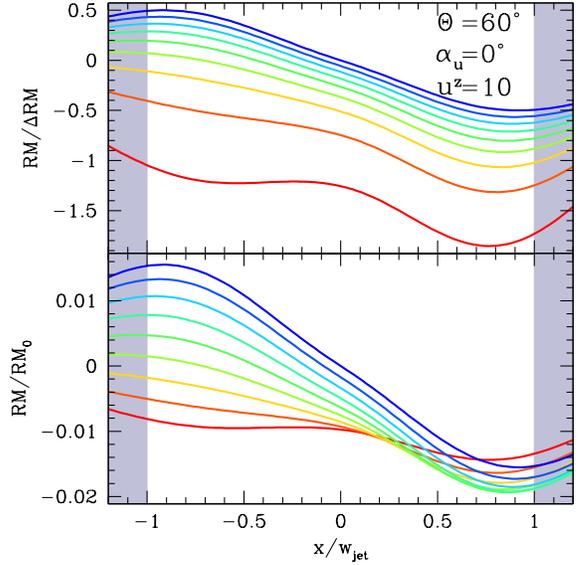}
\end{center}
\caption{Rotation measure profiles of an ultra-relativistic,
  purely outflowing, jet, $u^z=10$ and $\alpha_u=0^\circ$, as viewed 
  at $\Theta=60^\circ$, for magnetic pitch angle, $\alpha_b$, ranging from $10^\circ$ (red)
  to $90^\circ$ (blue) in steps of $10^\circ$.  Top and bottom panels
  are defined as described in Figure \ref{fig:static}.} \label{fig:vertical}
\end{figure}

\begin{figure*}
\begin{center}
\begin{tabular}{lr}
\includegraphics[width=0.43\textwidth]{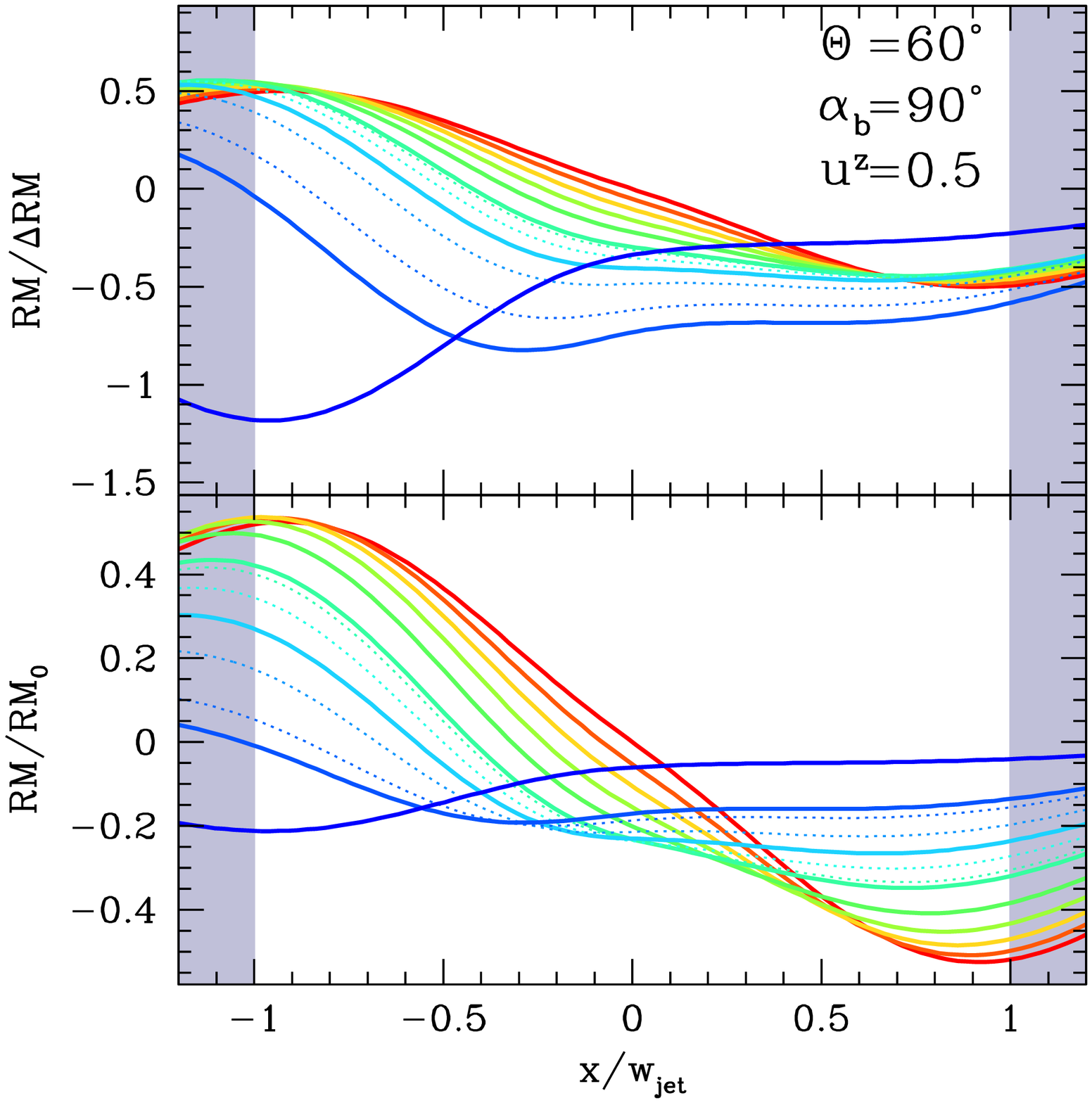}
&
\includegraphics[width=0.43\textwidth]{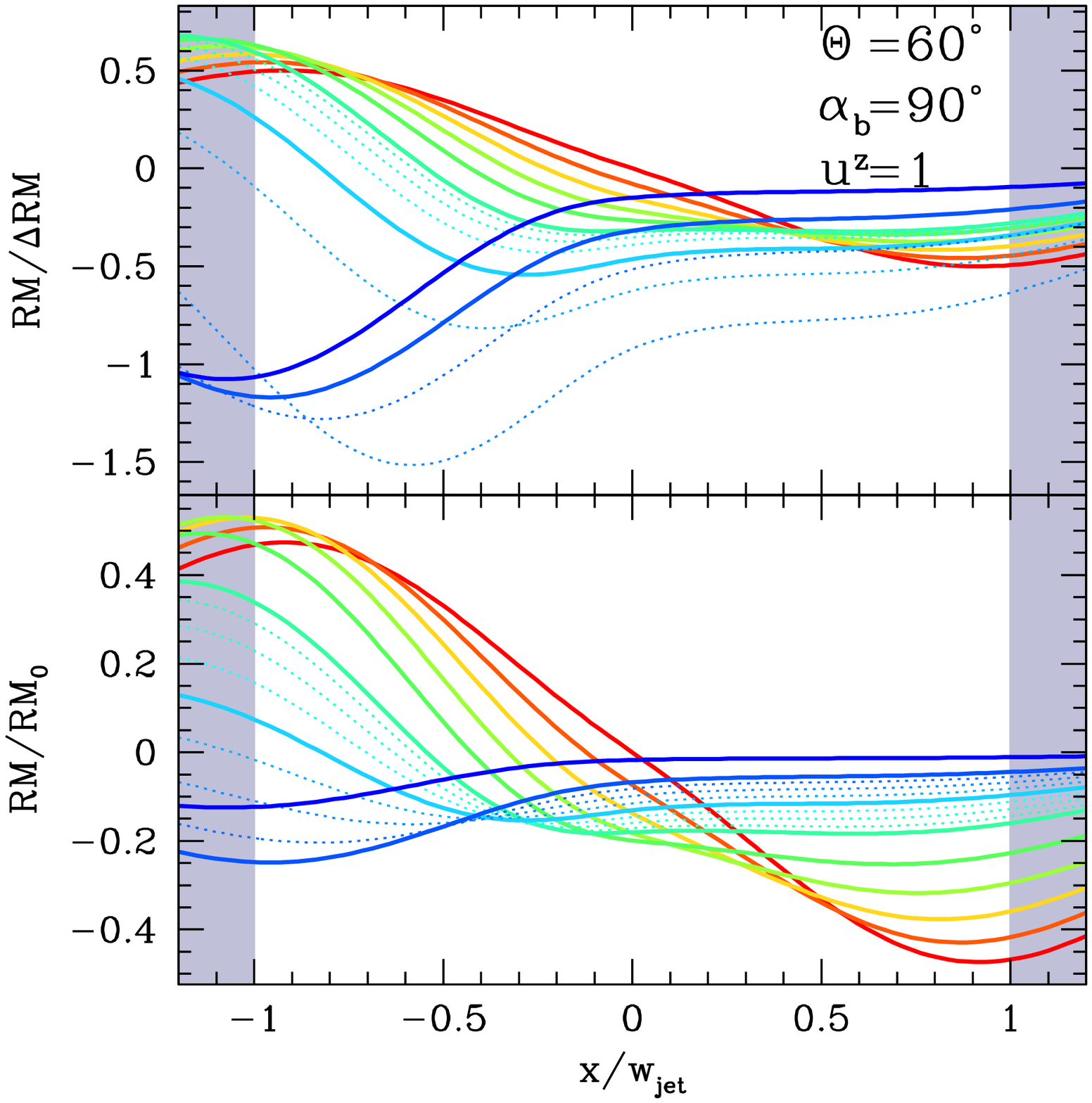}\\
\includegraphics[width=0.43\textwidth]{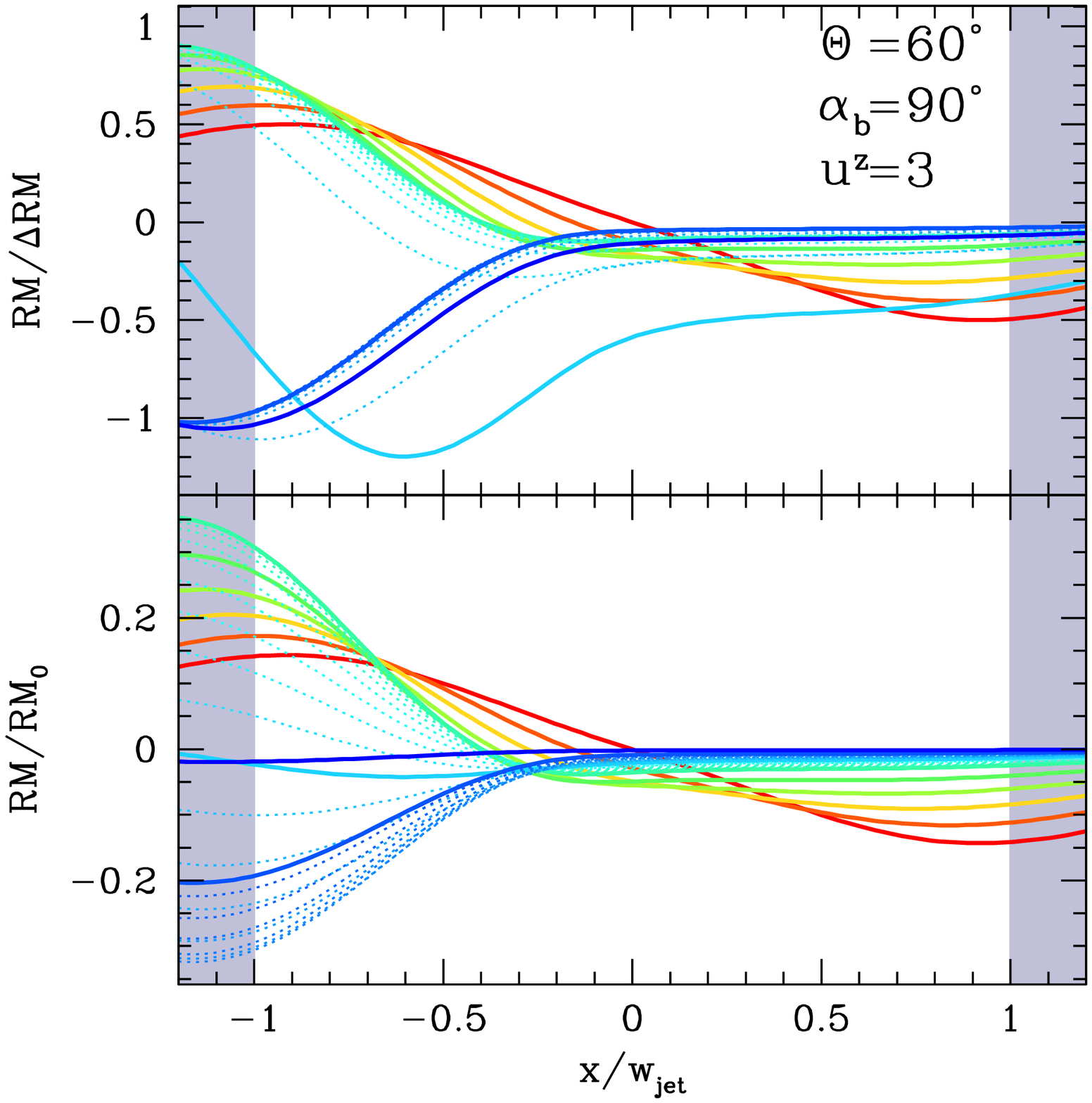}
&
\includegraphics[width=0.43\textwidth]{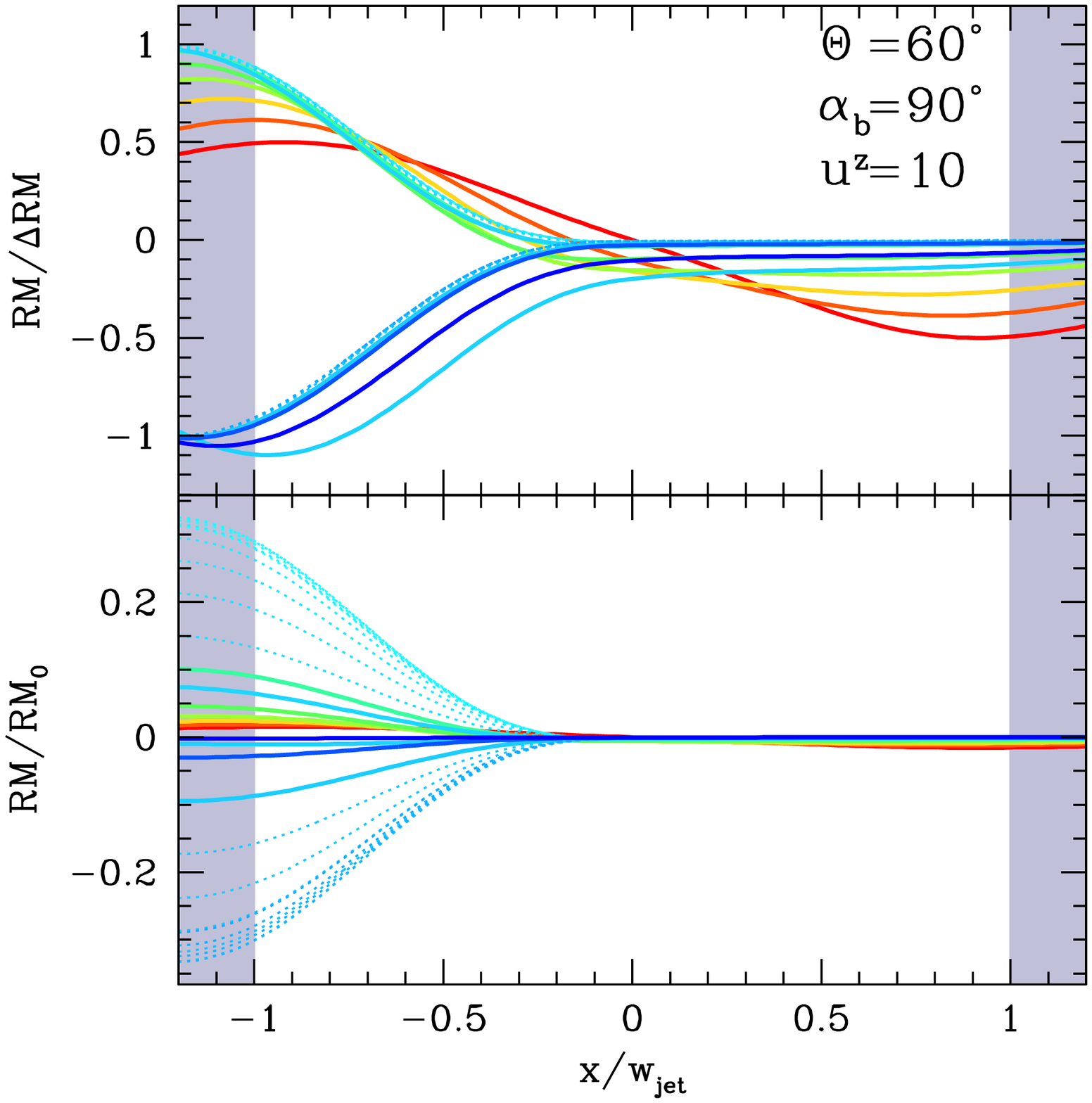}
\end{tabular}
\end{center}
\caption{Rotation measure profiles of jets with helical velocity
  fields for $u^z=0.5$ ({\em top left}), $u^z=1$ ({\em top right}),
  $u^z=3$ ({\em bottom left}) and $u^z=10$ ({\em bottom right}).  In
  all cases, $\alpha_b=90^\circ$,
  $\Theta=60^\circ$ and $\alpha_u$ ranges from $0^\circ$ (red) to
  $80^\circ$ (blue) in steps of $10^\circ$ for the solid lines.
  Dotted lines show the $\RM$ profiles for $\alpha_u$ distributed in
  steps of $0.1/\gamma$ about $\Theta$, highlighting the transition
  due to relativistic aberration.  Top and bottom panels
  are defined as described in Figure \ref{fig:static}.  We choose to
  fix $u^z$, and not $\gamma$, since $u^z$ is directly measured for
  the jet core by observations of superluminal motion.
} \label{fig:helical} 
\end{figure*}

Jet sheaths with relativistic bulk motion along the jet axis, but
without any helical motion, are similar to their static counterparts.
In this case the primary consequence of the motion for the $\RM$
profiles, apart from the drastic reduction in the net $\RM$, is
relativistic aberration, which effectively results in observers at
oblique angles viewing the jet from behind.  This is seen explicitly
for an extreme case in Figure \ref{fig:vertical}, in which the $\RM$
profiles are nearly identical to those for the static jet though
rotated $180^\circ$.  Most importantly, the sense of the $\RM$
gradient does not reverse as a consequence of the bulk
motion\footnote{For bulk motion along the jet to reverse the $\RM$
  gradient would require 
  $\Theta\gtrsim\gamma^{-1}$ while $\alpha_b\lesssim\gamma^{-1}$.
  While the former is almost certainly true, the latter would require
  magnetic field geometries that are nearly parallel to the jet
  axis.}.  The $\RM$ profiles do not significantly differ from the
case shown for $\gamma>3$, and for smaller velocities become even more
degenerate in $\alpha_b$.  Therefore, generally, within the context of axially
symmetric jet models, magnetic fields with moderate pitch angles are
incapable of producing strong asymmetric features in the $\RM$
profiles, though $\alpha_b$ and jet Lorentz factor do play a substantial
role in determining the absolute $\RM$.

In stark contrast, even moderately relativistic helical motion easily
produces dramatic asymmetric features in $\RM$ profiles, independent
of the magnetic field pitch angle, and in some cases reversing the
sense of the $\RM$ gradient.  As seen in Figure \ref{fig:helical},
this occurs both for trans-relativistic and ultra-relativistic jets,
and for velocity pitch angles as low as $30^\circ$.  Generic of
relativistic helical motion is the noticeable bow in the $\RM$
profiles.  This feature is only weakly dependent upon $u^z$ and
present for $\alpha_u\gtrsim30^\circ$ for $u^z\gtrsim1$.  Unlike
similar features for non-rotating jets, this occurs for a wide range
of absolute rotation measures.

When the revolving plasma in the jet sheath is approaching almost
directly, i.e., $\bmath{\beta}$ is within an angle $\gamma^{-1}$ of
$\bmath{\hat{k}}$, the direction in which relativistic aberration
rotates the magnetic field rapidly changes (see the dotted lines in
Figure \ref{fig:helical}).  As a result, the $\RM$ evolves
rapidly with $\alpha_u$.  At the same time, the Doppler shift reaches
its maximum, producing dramatic enhancements in the absolute $\RM$,
and for $\alpha_u$ very close to $\Theta$, complex $\RM$ profiles.
However, for all but the slowest jets, this is restricted to such a
small regime in viewing angle that it is relatively unlikely to be
observed in practice.  A much more important consequence is the
reversal of the sign of the $\RM$ profile gradient at
$\alpha_u=\Theta$.  This is seen most clearly in the bottom two panels
of Figure \ref{fig:helical}, in which the $\RM$ profiles associated
with velocity fields with high pitch angles (blue) are nearly
vertical reflections of those with low pitch angles (red).  Thus, not
only the symmetry of the $\RM$ profile is dependent upon $u^\phi$, but
also the direction of the $\RM$ gradient itself.

\section{Discussion} \label{sec:D}

\begin{figure}
\begin{center}
\includegraphics[width=0.43\textwidth]{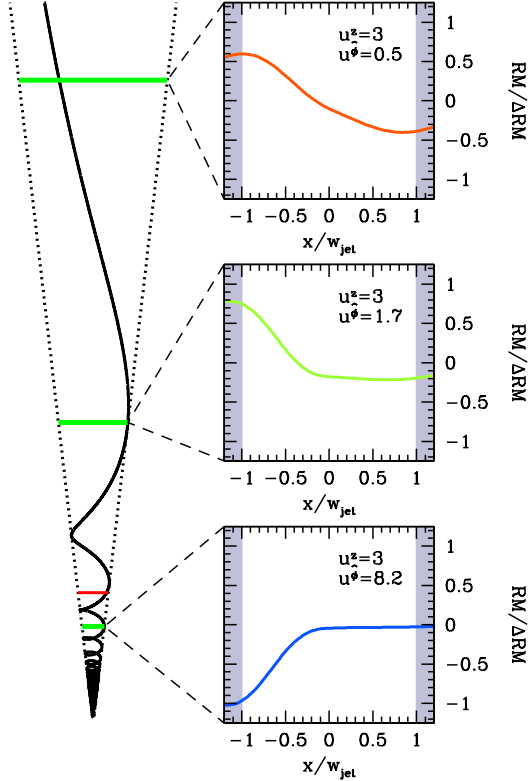}
\end{center}
\caption{Rotation measure profiles of an evolving jet.  The dotted
  lines show the envelope of the Faraday rotating sheath while the
  helical solid line is the path taken by a particle as it propagates
  within the sheath.  Throughout the jet $u^z=3$.  Below the red bar,
  $\alpha_u>\Theta$ (again taken to be $60^\circ$), while above
 $\alpha_u<\Theta$.  Positions along the jet where the $\RM$ profiles are
  shown at the right are marked by green bars.  The $\RM$ profiles are
  colored as in Figure \ref{fig:helical}.  Since the
  variation of the $\RM$ is dependent upon the magnetic field strength
  and plasma density in the jet, only the ratio $\RM/\Delta\RM$ is
  shown.  In each plot of the profile, the components of the
  four-velocity parallel and transverse to the jet axis are listed.
}\label{fig:jet}
\end{figure}

The presence of relativistic helical motion in the Faraday rotating
sheaths of AGN jets has a variety of specific and robust observational
consequences, some of which have already been observed.  These include
the presence of significant, frequency dependent asymmetric features
in the $\RM$ profiles, and even reversals in the $\RM$ gradient.
These are illustrated for a simple jet sheath model in Figure
\ref{fig:jet}.  In this case, $u^z$ is constant along the jet while
$u^\phi$, and thus $\alpha_u$, decreases with height.  Low
frequencies probe regions far from the AGN core
\citep[see, e.g.,][]{Blan-Koni:79}, where the transverse 
velocities are low and $\RM$ profile is approximately linear.
Conversely, intermediate frequencies probe regions closer to the AGN
core, where the transverse velocities are higher, which for even
modestly relativistic motions results in the characteristic bent
$\RM$ profile.  Sufficiently high frequencies may probe regions for
which $\alpha_u<\Theta$, if they exist, and thus exhibit reversed
$\RM$ profiles.

\citet{OSul-Gabu:09} present $\RM$ profiles for six blazars, as
measured in a variety of frequency bands.  In at least two cases
the $\RM$ profiles display the bent shape characteristic of helical
motion: 1156+295 and 1418+546.  In 1418+546 this is seen clearly in
a $\RM$ profile constructed from observations between $4.6\,\GHz$ and
$8.9\,\GHz$.  At higher frequencies, the $\RM$ profile shows evidence
of being reversed, though the absolute $\RM$ is reduced by nearly two
orders of magnitude, again consistent with the presence of
relativistic motion.  In contrast, at $4.6$--$8.9\,\GHz$, 1156+295
exhibits a nearly linear $\RM$ profile, while at $7.9$--$15.4\,\GHz$,
the $\RM$ profile becomes dramatically bent.  At higher frequencies
($12.9$--$43\,\GHz$), it becomes roughly linear again, which may be
evidence of reduced $u^z$ at small radii, a particularly fortuitous
view of the transition to a reversed profile, or failure to fully
resolve the $\RM$ gradient.  In addition 0954+658 and 2200+420
present dramatic changes in the absolute $\RM$, including its overall
sign, with changing observing frequency interval and radial distance
from the AGN core.  These are consistent with the observation of
regions transitioning between ultra-relativistic and moderately
relativistic helical motion close to the AGN core.

$\RM$ observations of knots in 3C 273 by \citet{Asad_etal:08a} provide
extraordinary examples of the bent profiles characteristic of helical
motion (cf. Figure 4 of that paper).  More interestingly, the $\RM$
profiles have evolved from 1995 to 2002, with the absolute $\RM$
increasing by roughly 40\% and the location of the bend in the profile
moving towards the jet center.  This behavior is consistent with
moderate slowing of the sheath during this period, i.e., a decrease in
the sheath velocities of roughly 20\% as they propagated outwards.

Unfortunately, most jet $\RM$ observations only marginally resolve
the transverse jet structure.  As a consequence it may be premature to
distinguish between Faraday rotation in nearby, though unrelated, gas clouds
\citep[see, e.g.,][]{Lain-Brid-Parm-Murg:08} and the ordered, helical
sheath models discussed here.  We note that 3C 273 is an exception,
{\em is} clearly resolved and provides a striking agreement with the
profiles we present.
However, space-based VLBI \citep[VSOP-2][]{VSOP:09} and millimeter-VLBI 
\citep[e.g.,][]{Brod-Loeb:09,Doel_etal:09} both promise order of
magnitude improvements in resolution, and have the potential to
resolve the nature of AGN jet Faraday screens conclusively.

\section*{Acknowledgments}
We thank an anonymous referee for helpful comments. This work was
supported in part by NSF grant AST-0907890.

\bibliography{jetrm}

\begin{thebibliography}{34}
\expandafter\ifx\csname natexlab\endcsname\relax\def\natexlab#1{#1}\fi

\bibitem[{{Asada} {et~al.}(2008{\natexlab{a}}){Asada}, {Inoue}, {Kameno}, \&
  {Nagai}}]{Asad_etal:08a}
{Asada}, K., {Inoue}, M., {Kameno}, S., \& {Nagai}, H. 2008{\natexlab{a}},
  \apj, 675, 79

\bibitem[{{Asada} {et~al.}(2008{\natexlab{b}}){Asada}, {Inoue}, {Nakamura},
  {Kameno}, \& {Nagai}}]{Asad_etal:08b}
{Asada}, K., {Inoue}, M., {Nakamura}, M., {Kameno}, S., \& {Nagai}, H.
  2008{\natexlab{b}}, \apj, 682, 798

\bibitem[{{Blandford} \& {Konigl}(1979)}]{Blan-Koni:79}
{Blandford}, R.~D. \& {Konigl}, A. 1979, \apj, 232, 34

\bibitem[{{Broderick} \& {Blandford}(2009)}]{Brod-Blan:09}
{Broderick}, A.~E. \& {Blandford}, R.~D. 2009, \apj, {\em in preparation}

\bibitem[{{Broderick} \& {Loeb}(2009)}]{Brod-Loeb:09}
{Broderick}, A.~E. \& {Loeb}, A. 2009, \apj, 697, 1164

\bibitem[{{Carilli} \& {Taylor}(2002)}]{Cari-Tayl:02}
{Carilli}, C.~L. \& {Taylor}, G.~B. 2002, \araa, 40, 319

\bibitem[{{De Villiers} {et~al.}(2005){De Villiers}, {Hawley}, {Krolik}, \&
  {Hirose}}]{DeVi-Hawl-Krol-Hiro:05}
{De Villiers}, J.-P., {Hawley}, J.~F., {Krolik}, J.~H., \& {Hirose}, S. 2005,
  \apj, 620, 878

\bibitem[{{Doeleman} {et~al.}(2009){Doeleman}, {Fish}, {Broderick}, {Loeb}, \&
  {Rogers}}]{Doel_etal:09}
{Doeleman}, S.~S., {Fish}, V.~L., {Broderick}, A.~E., {Loeb}, A., \& {Rogers},
  A.~E.~E. 2009, \apj, 695, 59

\bibitem[{{Ekers} {et~al.}(1969){Ekers}, {Lequeux}, {Moffet}, \&
  {Seielstad}}]{Eker-Lequ-Moff-Seie:69}
{Ekers}, R.~D., {Lequeux}, J., {Moffet}, A.~T., \& {Seielstad}, G.~A. 1969,
  \apjl, 156, L21+

\bibitem[{{En{\ss}lin}(2003)}]{Enss:03}
{En{\ss}lin}, T.~A. 2003, \aap, 401, 499

\bibitem[{{Hagiwara} {et~al.}(2009){Hagiwara}, {Fomalont}, {Tsuboi}, \&
  {Yasuhiro}}]{VSOP:09}
{Hagiwara}, Y., {Fomalont}, E., {Tsuboi}, M., \& {Yasuhiro}, M., eds. 2009,
  ASPC Series, Vol. 402, {Approaching Micro-Arcsecond Resolution with VSOP-2:
  Astrophysics and Technologies}

\bibitem[{{Hall} \& {Shukla}(2005)}]{Hall-Shuk:05}
{Hall}, J.~O. \& {Shukla}, P.~K. 2005, Physics of Plasmas, 12, 084507

\bibitem[{{Jones} {et~al.}(1996)}]{Jone_etal:96}
{Jones}, D.~L. {et~al.} 1996, \apjl, 466, L63+

\bibitem[{{Kharb} {et~al.}(2009){Kharb}, {Gabuzda}, {O'Dea}, {Shastri}, \&
  {Baum}}]{Khar_etal:09}
{Kharb}, P., {Gabuzda}, D.~C., {O'Dea}, C.~P., {Shastri}, P., \& {Baum}, S.~A.
  2009, \apj, 694, 1485

\bibitem[{{Komissarov} {et~al.}(2007){Komissarov}, {Barkov}, {Vlahakis}, \&
  {K{\"o}nigl}}]{Komi-Bark-Vlah-Koni:07}
{Komissarov}, S.~S., {Barkov}, M.~V., {Vlahakis}, N., \& {K{\"o}nigl}, A. 2007,
  \mnras, 380, 51

\bibitem[{{Laing} {et~al.}(2008){Laing}, {Bridle}, {Parma}, \&
  {Murgia}}]{Lain-Brid-Parm-Murg:08}
{Laing}, R.~A., {Bridle}, A.~H., {Parma}, P., \& {Murgia}, M. 2008, \mnras,
  391, 521

\bibitem[{{Lister} {et~al.}(1998){Lister}, {Marscher}, \&
  {Gear}}]{List-Mars-Gear:98}
{Lister}, M.~L., {Marscher}, A.~P., \& {Gear}, W.~K. 1998, \apj, 504, 702

\bibitem[{{L{\'o}pez}(2006)}]{Lope:06}
{L{\'o}pez}, E.~D. 2006, \apj, 641, 710

\bibitem[{{Lyutikov} {et~al.}(2005){Lyutikov}, {Pariev}, \&
  {Gabuzda}}]{Lyut-Pari-Gabu:05}
{Lyutikov}, M., {Pariev}, V.~I., \& {Gabuzda}, D.~C. 2005, \mnras, 360, 869

\bibitem[{{McKinney}(2006)}]{McKi:06}
{McKinney}, J.~C. 2006, \mnras, 368, 1561

\bibitem[{{McKinney} \& {Blandford}(2009)}]{McKi-Blan:09}
{McKinney}, J.~C. \& {Blandford}, R.~D. 2009, \mnras, 394, L126

\bibitem[{{Men} {et~al.}(2008){Men}, {Ferri{\`e}re}, \&
  {Han}}]{Men-Ferr-Han:08}
{Men}, H., {Ferri{\`e}re}, K., \& {Han}, J.~L. 2008, \aap, 486, 819

\bibitem[{{Noutsos} {et~al.}(2008){Noutsos}, {Johnston}, {Kramer}, \&
  {Karastergiou}}]{Nout_etal:08}
{Noutsos}, A., {Johnston}, S., {Kramer}, M., \& {Karastergiou}, A. 2008,
  \mnras, 386, 1881

\bibitem[{{O'Sullivan} \& {Gabuzda}(2009)}]{OSul-Gabu:09}
{O'Sullivan}, S.~P. \& {Gabuzda}, D.~C. 2009, \mnras, 393, 429

\bibitem[{{Taylor}(1998)}]{Tayl:98}
{Taylor}, G.~B. 1998, \apj, 506, 637

\bibitem[{{Taylor}(2000)}]{Tayl:00}
{Taylor}, G.~B. 2000, \apj, 533, 95

\bibitem[{{Taylor} \& {Perley}(1992)}]{Tayl-Perl:92}
{Taylor}, G.~B. \& {Perley}, R.~A. 1992, \aap, 262, 417

\bibitem[{{Tchekhovskoy} {et~al.}(2008){Tchekhovskoy}, {McKinney}, \&
  {Narayan}}]{Tche-McKi-Nara:08}
{Tchekhovskoy}, A., {McKinney}, J.~C., \& {Narayan}, R. 2008, \mnras, 388, 551

\bibitem[{{Vlahakis} \& {K{\"o}nigl}(2004)}]{Vlah-Koni:04}
{Vlahakis}, N. \& {K{\"o}nigl}, A. 2004, \apj, 605, 656

\bibitem[{{Walker} {et~al.}(2000)}]{Walk_etal:00}
{Walker}, R.~C. {et~al.} 2000, \apj, 530, 233

\bibitem[{{Zavala} \& {Taylor}(2001)}]{Zava-Tayl:01}
{Zavala}, R.~T. \& {Taylor}, G.~B. 2001, \apjl, 550, L147

\bibitem[{{Zavala} \& {Taylor}(2002)}]{Zava-Tayl:02}
{Zavala}, R.~T. \& {Taylor}, G.~B. 2002, \apjl, 566, L9

\bibitem[{{Zavala} \& {Taylor}(2003)}]{Zava-Tayl:03}
{Zavala}, R.~T. \& {Taylor}, G.~B. 2003, \apj, 589, 126

\bibitem[{{Zavala} \& {Taylor}(2004)}]{Zava-Tayl:04}
{Zavala}, R.~T. \& {Taylor}, G.~B. 2004, \apj, 612, 749

\end{thebibliography}

\end{document}